Spatial, Social and Data Gaps in On-Demand Mobility Services: Towards a Supply-Oriented MaaS

Ronit Purian[1] and Daniel Polani[2]

**Abstract**

After a decade of on-demand mobility services that change spatial behaviors in metropolitan areas, the Shared Autonomous Vehicle (SAV) service is expected to increase traffic congestion and unequal access to transport services. A paradigm of scheduled supply that is *aware of demand but not on-demand* is proposed, introducing coordination and social and behavioral understanding, urban cognition and empowerment of agents, into a novel informational framework. Daily routines and other patterns of spatial behaviors outline a fundamental demand layer in a supply-oriented paradigm that captures urban dynamics and spatial-temporal behaviors, mostly in groups. Rather than real-time requests and instant responses that reward unplanned actions, and beyond just reservation of travels in timetables, the intention is to capture mobility flows in scheduled travels along the day considering time of day, places, passengers etc. Regulating goal-directed behaviors and caring for service resources and the overall system welfare is proposed to minimize uncertainty, considering the capacity of mobility interactions to hold value, i.e., *Motility* as a Service (MaaS). The principal-agent problem in the smart city is a problem of collective action among service providers and users that create expectations based on previous actions and reactions in mutual systems. Planned behavior that accounts for service coordination is expected to stabilize excessive rides and traffic load, and to induce a cognitive gain, thus balancing information load and facilitating cognitive effort.

**1. Introduction**

The fast development of mobility services creates new patterns of spatial behaviors. While the location-based applications on our cellular phones offer mobility services that could design sustainable mobility patterns, the current ecosystems of smart mobility service providers and autonomous vehicles (AV) are far from environmental objectives and the social demand for equitable access to transportation in diversified communities (e.g., ITF, 2021; Rodier, Jaller, Pourrahmani, Pahwa, Bischoff, & Freedman, 2020; WEF, 2019). Moreover, navigation applications and especially the broadening implementation of autonomous vehicles are expected to increase traffic loads and unequal access to mobility services (e.g., Ben-Dor et al., 2019; Vazifeh et al., 2018).

---

[1] Tel Aviv University https://orcid.org/0000-0002-8889-7136
[2] University of Hertfordshire https://orcid.org/0000-0002-3233-5847

A better coordinated group ridesharing is vital to stabilizing traffic loads, optimizing freedom of movement and delivering equitable access to mobility services (Kaddoura, Bischoff, & Nagel, 2020; Narayanan, Chaniotakis, & Antoniou, 2020; Vosooghi, Puchinger, Jankovic, & Vouillon, 2019).

The paper conceptualizes – from both the social sciences and urban studies perspectives – a platform for *group* ridesharing, which addresses several challenges. First, the current lack of mobility data sharing severely limits the capability of both suppliers and demanders to envision and create an effective mobility ecosystem. Approaches to automatic coordination in the domain of group ridesharing in order to enable flexible traffic flows, based data sharing from peers, should be constructed as a decentralized system from different data sources. Second, data frameworks to mobility require extensions to include appropriate parameters of spatial behaviors and collective urban dynamics, externally validated models (Al Sayed, Bew, Penn, Palmer, & Broyd, 2015; Boeing, 2021; Rodrigue, 2020, p. 177, 217), with the ability to link cognitive agent and multiagent (van Dijk & Polani 2011, 2013; Harder et al., 2010) and collective models of spatial cognition, urban- and regional-level interventions, and observed responses.

Third, the ability to properly assemble mobility data and to model spatial behaviors requires a systematic understanding of the relationships between the different social groups, properties of their mobility, and their respective vulnerabilities and preferences (e.g., WEF, 2019).

The overall objective of the paper is to support the construction of an operative mobility platform, introducing coordination and social and behavioral understanding, including urban cognition and empowerment of agents, into a novel informational framework to facilitate service design. The principal-agent problem in the smart city is a problem of collective action among service providers and users that create expectations based on previous actions and reactions in mutual systems (Bak-Coleman et al., 2021; Marx, 1859; Weyl & White, 2014).

Recent studies have been setting individual cognition and the research of wider social attitudes closely together (Jost, Halperin & Laurin, 2020; Woelfert & Kunst, 2020; Zmigrod et al., 2021) and the design and implementation of smart services joins them in closer association along urban problems and challenges such as uncertainty, information load, social discrimination, institutional trust, aversion of interactions with other riders, and urban scaling and marketization that put limits on relationship within local communities that are based on mutual understanding. Geography had introduced spatial thinking to the analysis of human behaviors. As it is rooted in the spatial dynamics of groups, communities, and individuals in groups, a review of the literature is beneficial to better understanding research gaps and to realize which opportunities are available and have the potential to provide possible solutions.

## 2. Service design through social inquiry

The MaaS anomaly erupts as more and more travels are carried out in an individual on-demand mode of service. Transportation Network Companies (TNC) such as Uber and Lyft are associated with increased traffic load (Clewlow & Mishra, 2017; Diao, Kong, & Zhao, 2021; Erhardt, Roy, Cooper, Sana, Chen, & Castiglione, 2019; Hall, Palsson, & Price, 2018). With the introduction of the Shared Autonomous Vehicle (SAV), much as happen with TNC, simulations demonstrate increase in traffic volume in different pricing schemes, including congestion charge (Kaddoura, Bischoff, & Nagel, 2020). Congestion charge is often proposed to prevent excessive use of mobility resources, however, the effect of congestion charge may not suffice to reduce congestion (ITF & OECD, 2021). According to a simulation of various operation costs and pricing setups, it appears that SAV operator profit and user trip-specific benefits and costs are not enough. Congestion fees may even result in a lower system welfare if charged only from SAV users and not from users of conventional cars, as SAV users would tend to take a conventional car instead, to avoid the fee. If the goal is to reduce the number of vehicles on road, then pricing, routing and vehicle dispatch strategies are the setting that account for the overall system welfare (Kaddoura, Bischoff, & Nagel, 2020). Pricing and vehicle routing and dispatch strategies offer a crucial advantage for the transport system through their impact on user choice. Decisions of transport users on the mode of transportation depend on effects of demand and supply (Lamotte, de Palma, & Geroliminis, 2017; Mladenović, & Haavisto, 2021), that are intrinsically connected to the configuration of operation and service specification, e.g., critical fleet size, optimal vehicle occupancy and pricing scheme, vehicle-redistribution (collecting vehicles from low demand areas and repositioning in high demand areas to rebalance supply), range of charging stations and maximum driven distance, and access to real-time data, being integrated in multimodal transportation network (Ben-Dor et al., 2019; Vosooghi, Puchinger, Jankovic, & Vouillon, 2019). Small fleets that manage to address the demand in peak-hour perform well when responding to *dynamic* demand, i.e., given the availability of real-time (and not static) demand data. In addition, user choice depends on attitudes and demographic. The effect of gender, age, profession and household income on attitudes and trust is revealed through the willingness of travelers to choose SAV and the perceived utility of SAV, based on travelers score, in similar trips – across the same conditions of time and cost (Vosooghi et al., 2019). However, change in user expectations reframes attitudes, and therefore affects mobility decisions, and actual spatial behaviors.

Service models that preserve a status quo of personalized supply, provided on-demand, have direct and indirect effects. While direct indicators of road safety, air pollution, and time spent in traffic jams make the costs of traffic congestion observable and quantifiable, other influences are invisible. Quantifying the impact of SAV services had already raised the need in policy intervention (Narayanan et al., 2020; Rodier et al., 2020). Furthermore, the hidden interactions among social, environmental, and economic effects of traffic congestion may have unexpected consequences, as spatial behavior depends on values and preferences that are subtle.

**Interfaces among the human, the habitat, and the IT artifact**

Current studies show the anomaly of mobility services. The use of information technologies (IT) has already proved to *increase the number of trips* while tapping into basic needs of travelers, and changing destinations for existing trips to more circular routs (Dal Fiore et al., 2014; and Figure 1). Trips to and from a point of gravity, connecting center to outer edge, change to more circular routs, as if central locations lost gravity in the age of ride-hailing services.

Design service lays in the complex interplay among the human, the habitat and the IT artifact. The IT artifact, new service tools entangled with many agencies, call for a living theory in smart services and the capacity of mobility to hold value, the *mobility capital* – termed *motility* – through mobility choices of individuals and as a policy objective (Shliselberg & Givoni, 2018; 2019; Shliselberg,

Givoni & Kaplan, 2020; Noy & Givoni, 2018). Every few months a new mobility service comes out, and we need several apps to plan a trip, book a ride, pay and travel, walk and navigate. The information load increases, as well as population density in our immediate vicinity. To avoid the risk of modal shift from public transport (PT) to ridesharing services, mobility services are expected to be integrated in a multimodal transportation network, e.g., with SAV services complementary to PT for the last mile, or in low demand areas. However, service providers such as Uber and Lyft operate an independent mode of service (Narayanan, Chaniotakis, & Antoniou, 2020). The desired fusion of mobility services depends on their coordination with an efficient public transport system, i.e., *coordination on the supply-side*; however, main mobility services today adjust to demand. An organizing vision is absent. Smart utilization of available supply could facilitate congestion and reduce parking distress that large fleets are facing during low-demand hours. There is a need in coordination among stakeholders, including users, to utilize available supply. But, rather than balancing, the network structure of the service economy incentivizes the domination of few players over unorganized masses of individual users. The lack of organizing themes appears in several service levels. Services allow users to share products but, in the current form of sharing economy, users refuse to share their space and time with others, thus stimulating excessive use in the case of mobility services. Public transportation is subsidized in rural areas to guarantee equal access to mobility services; however, although vehicles are very often empty and demand is low, smart services are rarely applied to schedule rides and to adjust vehicles to clusters of demand.

To present a clear problem definition, Figure 1 shows the impact of technology on mobility patterns, respectively integrating conceptions of mobility in a historical overview to reflect altering perspectives of time, space, and the social interaction.

*Figure 1. How we move in cities: The sociotechnical patterns of Mobility as a Social action*

A historical perspective on Mobility as the *settings of groups* (Hägerstrand, 1970); as the *processing of procedures* (Axhausen, 1995); and *transposing of functionality* (Aleta et al., (2020). Respectively, excessive rides (A-C) increase the number of routes (Dal Fiore et al., 2014) and centrality lost gravity. In a separate paper we present the underlying assumptions that derive the guiding principles towards mobility coordination that is aware of demand but not on-demand. In this paper we focus on the reasoning of scheduled supply (Rethinking Group Ridesharing for Community Shared Economy).

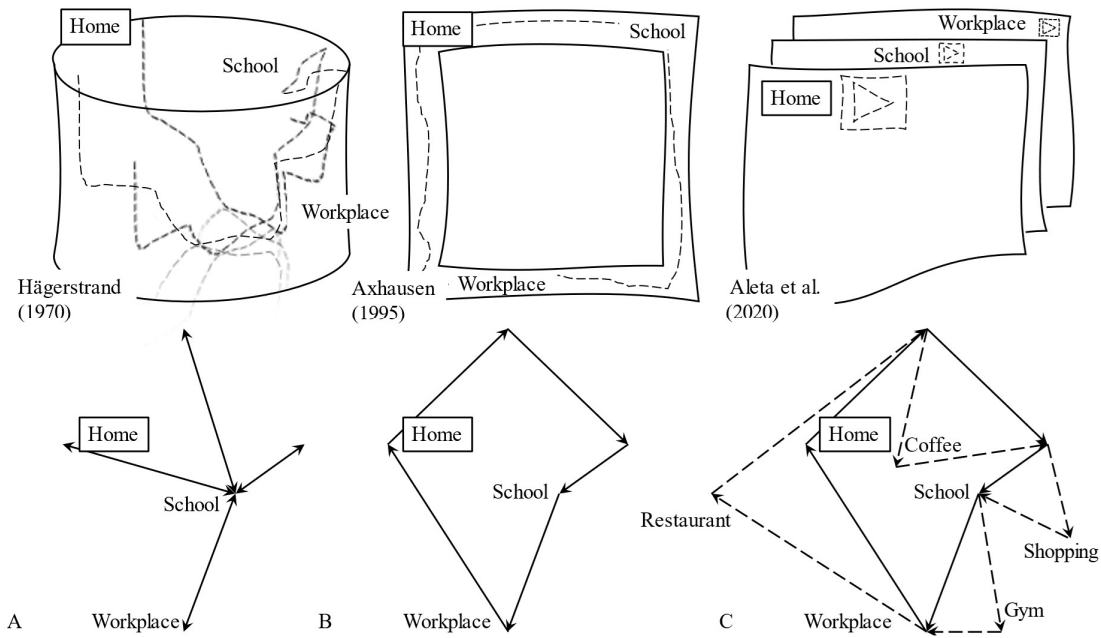

Shared-mobility studies are mostly about on-demand services rather than scenarios with planned, predefined requests, although "reservation-based systems enable better planning of routes and schedules and, if optimally designed, higher efficiency, thereby reducing fleet size, empty cruising time and operating cost" (Narayanan et al., 2020; p. 4). With the focus of most research on responsive service, several knowledge gaps exist regarding resource utilization, SAV operations, and even traveler behavior and responses of individuals (Narayanan et al., 2020). Reservation-based services suit different types of trips, and have the potential to stabilize demand and supply across areas, weekdays and peak hours, balancing excessive rides in high-density areas and improving fleet distribution to provide equal access to mobility service. Narayanan et al. (2020; p. 25) conclude that although "a vast amount of research is being published every year, a considerable number of research gaps still exist […] resulting in most studies inevitably basing their analyses on strong assumptions. […] the current modelling methodologies and the analyses of impacts of SAVs need to include a set of scenarios based on plausible assumptions." By ensuring vehicle availability in predefined location and time, thus reducing uncertainty regarding the expected service, reservation-based systems have the potential to shift private car users to shared services. In addition, ridesharing in a group enables rebalancing that leads to shorter waiting time, while the individual ride scenario remains without the ability to rebalance (Narayanan et al., 2020). Redistribution of excess vehicles is important for economic viability and to provide equal access to mobility services.

Paris car-sharing service Autolib operated in a high-density region and yet failed (Bolloré Group, 2018, a, b) and emphasized the disadvantage of car- vs. ride- sharing scheme that allows a collaborative service design. The individual car sharing service failed to prove economic viability – mainly redistribution costs. repositioning vehicles by morning in metro stations in the inner-city centers, ready for commuters, after their evening trips, over multiple locations, on their way back home; and then again rebalancing outflow to Paris suburbs by evening, and so on; the undesirable anomaly of mobility services. In addition to redistribution costs, Paris slums were accused of failing the system. The potential market of the outer city ring became a target of blame for complaints about "dirty, poorly maintained cars" (Bloomberg, 2018; Straits Times, 2018).

When considering planning, combining pick-up and drop-off locations with walkable trajectories and accessibility to green places may contribute to user satisfaction – and furthermore help in building trust in a carefully planned service – and encourage richer system engagement, which is expected to induce public participation, not only for the purpose of public campaign and adoption rate but to better utilize the service system, relief privacy concerns and enhance data sharing and a collective action in mobility decisions (Ostrom, 2009; 2010a, 2010b; Said, Fischer & Anders, 2020; Woelfert & Kunst, 2020).

Therefore, trust and community building are not only enablers in the proposed mobility system, but an expected contribution of the ideal system. Participatory planning is part of the design of the proposed ridesharing service and its processes. In ridesharing or car-pooling (Shaheen & Cohen, 2019) the number of passengers increases travel time and length. Service configuration that specifies the operative decision to decline a request that increases time/length is a decision-rule to reject a person. Moreover, the overall system welfare depends on service configuration that affords pooling and ridesharing that increase the capacity of existing vehicles, reduce traffic loads, and improve performance (e.g., redistribution) compared to individual car-sharing or SAV services. While travel time and length increase for existing passengers, i.e., service gets worse – for future passengers the waiting time improves; and service availability increases for potential users. The whole system operates for the benefit of society at large and improves also environmental measures; however, to support the needs of a general population of passengers, the competition among conflicting preferences should be resolved. To what extent should MaaS consider the overall performance on the expense of individual needs, and how can design solve conflicting needs, e.g., providing a service in low-demand area? Can service design address the tension between main service measures?

Long detours are considered to indicate the fault or inadequacy of profit-driven service providers. According to Bischoff et al. (2018), a "high revenue ratio may not necessarily be good, as it can also mean that passengers are transported for long detours" (Bischoff et al., 2018; p. 14). However, flexibly providing passengers with the ability to choose/avoid routes; having the freedom to altruistically "contribute" time and extend a ride in order to accept specific requests; not limiting access to distant low-demand areas; and so on, are critical public-level considerations that shape a service design. At the same time, rich user engagement in the interactions with the system and other users, over multiple rides, could play a social role and allow agents to maximize their utility while also redefining their utility and reshape relationships with other users, processes, and system tasks and functions (Al-Natour & Benbasat, 2009), thus utilizing the system through richer system use (Burton-Jones & Straub, 2006). Accounting for behavioral change, in and between interactions, holds an operative rationale to mobility (and, in its turn, social implication). If aggregating passengers is a possible solution, then how to create a group, and what is the meaning of "community building" as a design goal, when planning the expected service.

**Community building: Groupness by design**

Urban routines provide opportunities to group people together, even when "waiting for a bus at a bus stop", as Sartre describes, to crystalize the instantiation of human groupness in 1960's "Critique of dialectical reason" (Sartre, 2004; p. 259). How to utilize the new quality that the *group* introduces to a collection of individuals is a design goal (Bak-Coleman et al., 2021). The whole is more than the sum of the parts, stated Herbert Simon, while reflecting on the "laws of their interaction" and the "properties of the whole" (Simon, 1969; p. 184). The instances of human groupness (Young, 1994) seem to reduce in cities that become increasingly crowded, wired and service-based (e.g., Supply Chain Trade and Workers in a Globalized World report, World Bank, 2019). Rather than connecting, the amalgamation of information systems into platforms and applications, a process that is closely related to urbanization and globalization, disconnects people, even in close proximity (Purian, 2020).

This emphasis on community building is a behavioral aspect to guide the design of the *socio*-technical system for group ridesharing, understanding the connection between digital and physical spaces, and the importance of place-making and belonging. Needs of belongingness create our communities – through friendship, trust, and acceptance (Maslow, 1958). The social fabric therefore provides a sense of connection that protects our identities. Being part of a group is one of the pillars in a theory of human needs and the case of digital presence, physical location, and narratives acquired both digitally and physically for work immigrants, tourists, commuters and local residents (Purian, 2020).

Ridesharing with other city dwellers and visitors is mostly developed in the domain of carpooling; however, sharing in mobility is not restricted to the type of cars and drivers, neither to application-based carpooling, but a service open to support ridesharing of any kind, including SAV services that were reviewed above. The social interaction involved in such services is a major barrier to the acceptance and use of mobility services in groups, as recent studies show.

**Discrimination and social aversion**

Disparities due to neighborhood demographics exist in the traditional taxi industry, and research shows how it can work both ways, i.e., taxi drivers miss earnings opportunities due to discrimination when they avoid neighborhoods (FRBB, 2018). Personal prejudice (termed taste-based discrimination) is discerned from unequal provision of taxi services due to incomplete information about the potential demand across neighborhoods for the taxi services (termed statistical discrimination, i.e., lack of knowledge, particularly for black and Asian residents). When drivers gain experience, the discrimination across neighborhoods decreases as well as their "wage uncertainty" (FRBB, 2018). The study concludes that unequal access to taxi services depends on wage anticipation. When serving different social groups, experienced taxi drivers can better predict wage variation, and adjust, i.e., choose the areas they serve according to income expectations.

To contradict this self-interested/instrumental conclusion, discrimination in the traditional taxi industry can be compared with discrimination by TNC drivers (Uber and Lyft). Evaluation of wait times and request cancellations found racial biases in both the taxis and the ride-hail services, however, discrimination was much stronger in traditional taxis: the probability of request cancellation for black compared to white riders was 73% for a taxi service, compared to 4% on both Uber and Lyft (Brown, 2018). A possible explanation would look at the drawback of the gig economy, rather than social progress and labor rights, as technology advances (EPI, 2018; Farrell, Fiona, & Amar, 2018). Indeed, for Uber drivers, workplace flexibility and technology advances in the industry of mobility services did not provide the expected remedy that may close the gender pay gap (Cook, Diamond, Hall, List, & Oyer, 2018). Women are left behind in the gig economy. The proposed reasons are not related to taste-based customer discrimination but to experience, and preferences such as choosing areas near home, safety concerns, and driving speed. Unlike Uber in-app GPS and fluctuating prices, taxi fares are fixed. Having incomplete information (statistical discrimination) may affect driver earnings. In addition, evidence from social science suggest a potential bias via the customer rating system, that may affect employment status (Rosenblat, Levy, Barocas, & Hwang, 2017).

Yet, MaaS provide a robust platform for discrimination against the customers as well (Ge, Knittel, MacKenzie & Zoepf, 2016; WEF, 2019). While the choice about serving a given neighborhood, a decision that takes place on the supply side in the labor market for taxi service, is a labor market discrimination that can affect employment or wages, such market outcomes also have social implications on the demand side of transport users. Discrimination is mostly studied based on age, ethnicity, gender, or race (Sarriera et al., 2017). Discrimination in response to people who request a wheelchair-accessible ride is illegal, yet people with disabilities are overlooked (NYLPI, 2018; 2019; SFMTA, 2019).

Longer waiting time, not even knowing whether a vehicle is available, is a growing component of uncertainty in urban life for residents in different social groups. The expected social interactions in transportation affect the decision of TNC users on the mode of transportation, whether to share a ride and pay less. When considering behavioral and social aspects, in addition to the time and cost trade-off (i.e., longer routes and travel time to pick up and drop off other passengers), the motivation to avoid a negative interaction is stronger than the incentive of a positive social interaction, as expected when studying risk aversion (Sarriera et al., 2017). This is generally valid for punishments and rewards. Punishment is feared more than reward coveted. Personal prejudice, in this case a discrimination based on social class and race, raised the need in information about other passengers (Sarriera et al., 2017). Prejudice is in a way a factor that induces uncertainty. Feeling unsafe, as reported by many women, leads to a gender bias, especially women preferring other women passengers.

When considering ease and speed, ridesharing was the preferred choice over walking and public transportation (Sarriera et al., 2017). Thus, after the longer travel time and the lower price, the social interaction with other riders was a negative influence that raised uncertainty and risk aversion in ridesharing. SAV could avoid the "driver-rider discrimination" that taxi and TNC services share, but not to prevent discrimination at all, and the promise to combine multiple trips into a single ride in a shared vehicle and reduce traffic congestion faces, according to recent studies, the "rider-rider discrimination" (Moody, Middleton, & Zhao, 2019). Discrimination in pooled rides reveals a social struggle, intensified by certain passenger characteristics that hold specific ethnic and class attitudes. When analyzing racial and socioeconomic biases, the outcomes show that "race does not have a significant effect on discriminatory attitudes, but white respondents that live in majority white counties are more likely to hold discriminatory attitudes with regard to race (no effect is observed regarding class preferences). The same is true of respondents that live in counties in which a larger share of the electorate voted for the Republican candidate in the 2016 presidential election. Conversely, higher-income respondents appear more likely to hold discriminatory attitudes regarding class, but no effect is observed regarding racial preferences" (Middleton & Zhao, 2020; p. 2391). This is a case where the so-called privileged, or empowered agent, repels its underprivileged opposite (e.g., upper class discrimination against lower class is more likely than race discrimination). Either class-based discriminatory attitudes or racial discrimination, the findings emphasize the social and the economic characteristics of higher-income respondents (although racial preferences were not observed, class had an effect) and the characteristics of counties with white majority (although class preferences had no effect, racial bias was observed) (Middleton & Zhao, 2020). While a neutral driver could perhaps supervise and play the role of a moderator among riders, the "rider-rider discrimination" in the driverless service requires – once again, now in a social context – an intervention in mobility services: either radicalizing segregation as already happens in women-only carriages around the world (e.g., Graham-Harrison, 2015) or, rather than separation, improving "rider-rider interactions" through service design, providing information and encouraging decisions, engagement and behavioral choices with interfaces and mechanisms that enhance civic mindedness and attitudes toward social action (Purian, 2019). While Middleton & Zhao (2020; p. 8) claim that, without a driver, "passengers will need to establish trust and accountability among one another", we propose that trust and accountability are a coherent value of the service-system, afforded to a broad variety of users and stakeholders, including passengers and drivers, and not a task administered for them to establish. A possible service goal is therefore to improve interactions through appropriate interventions. That way we form a data-knowledge-action system: organize the information and create new knowledge structures that lead to action and enhance agency and a sense of control. When we reveal or imagine hidden connections and structures, rather than maintaining a status quo in existing roads and routines, we obtain a system goal that is to reduce uncertainty, and an ultimate design goal to restore empowerment. *Coordination is needed at all levels*.

Urban cognition and movement in the built, crowded environment of cities have certain laws and features that interact with the image of the city. Regular mobility patterns affect the way we perceive the city, and movement in the city affects the representations created and transform while experiencing the city (Portugali & Haken, 2018). The implication of movement on mental processes is estimated in terms of information.

The framework provided by information theory and urban cognition to analyze and interpret urban behaviors aligns with the free energy principle that guides self-organizing systems in a way that strives to minimize the energy allocated for adaptions to changes. To stay at equilibrium while facing constant change, the free energy principle puts boundaries to *surprise*. An agent inhibiting irrelevant information, learning to avoid it, implicitly bounds the cognitive load of irrelevant information that is constantly being generated due to constantly changing environment (Friston, 2010; Haken & Portugali, 2021; Weiner et al., 1994). The interplay between change and the mental representation that struggles to adapt creates information load (or cognitive load), which is undesirable, therefore an event that generates new information or requires to reach a decision is considered a situation that imposes costs which agents seek to avoid.

**Urban dynamics and urban cognition**

Patterns in human behavior, and for the purpose of this paper – patterns in spatial behavior, can be assimilated in mechanisms of urban services. The affordance of behaviors by existing patterns is, in a sense, a design mechanism that enables empowerment; empowerment is all about affordances. It uses the embodiment, and the embodiment defines what affordances you have. From the perspective of urban cognition, a second parameter, the freedom of agents in the choice of their movement options, can be modeled using *empowerment*; the options available to an agent in a given situation (Salge, Glackin & Polani, 2014b). The free energy principle is about the constantly changing environment that forces agents to minimize cognitive load (Biehl, Guckelsberger, Salge, Smith, & Polani, 2018; Friston, 2010) rather than creating internal representations of urban places (Haken & Portugali, 2021).

Self-organization is a mechanism of autonomous coordination between individuals that has a spontaneous outcome. When the relationships among the individuals are defined, even by informal rules, a mix of planned and spontaneous actions and reactions is formed and further coordinates the evolving groups of individual entities.

Research today reveals layers of attitudes, preferences, and perceptions, perhaps hidden or subliminal, as reviewed above (social preferences, discrimination). The social schemes can be utilized to guide the design of service pipelines. In this context of patterns in spatial behavior, it should be emphasized that, although disorder emerges, some order is assumed to be desirable, practically to stable traffic and socially to enhance a meaningful urban rhythm with repeated ebb and flow which are tolerable, in predictable cycles of day and night, weeks and seasons. The social and environmental goal is to reduce uncertainty and unavailability of mobility services due to heavy demand and constant traffic jams. To gain this order, the *activity space* (Axhausen, 2007) of individuals and groups is postulated.

Although technology disrupted common routines and increased mobility, a person's daily trajectory still is assumed to preserve a *radius of gyration* that contains most routes and locations (Dal Fiore et al., 2014; Gonzalez et al., 2008; Psyllidis, 2016). But, taking into account commuting today and the way mobility services change mobility patterns (the case of Paris), are commuters endangered species? (Benenson & Ben-Elia, 2019; Zivion et al., 2017).

## 3. Model development

A novel *informational framework* in the context of smart mobility is proposed to complement the outlined approach to *urban behaviors*, modelling individual moving agents (whether walking, driving or using other means of mobility) decision-making based on information-theoretic measures of cognitive load. The objective is to provide the reasoning for further formulating cognitive multiagent models that capture plausible decision-making under bounded rationality assumptions and incomplete knowledge and information-processing capacities of the individual agents, in areas and geometries that address urban morphology, geographically characterized in structural and functional principles.

**Cognitive multiagent models.** Cognitive multiagent models informed by information-theoretic constraints; taking into account the cognitive load that will affect how human decisions are made, provide indications of typical population-level decision-making tendencies, such as seemingly irrational cognitive shortcuts, collective misconceptions, topology-induced location and strategy preferences. This will permit us to synthesize agent dynamics which is behaviorally plausible.

**Empowerment-Based Dynamics (empowerment models).** Methods for empowerment-generated behaviors for the multiagent systems can be used to model decision-making locally, under the assumption that these follow the empowerment maximization hypothesis: this assumes that natural agents aim to maximize the external information-theoretic channel capacity of their action-perception loop of agents (e.g., Klyubin et al., 2008; Salge et al., 2014a; 2014b). Importantly, the local information only pertains to individuals acting in a collective, including the compensatory behavior of individuals as reaction to an environmental change or regulatory constraint; and to stabilization and "calming" overall multiagent dynamics through absorption of external shockwaves impinging on the system, without requiring an explicit external centralized control mechanism (Everding, 2013).

Empowerment is integrated into the multiagent scenarios to investigate the effects of various modifications such as topology, structure, regulations, or other constraints on the mobility. Specifically, consequence of top-down imposed modifications are at focus, investigating local decision-making as the "evasive" reactions of the agent collective to the imposed conditions. This leverages the empowerment-induced tendency to explore the more "extreme" astatistical types of behavior (which is more plausible than statistical/typical behavior for natural cognitive actors). At the same time, due to the "bubble of autonomy" that empowerment induces around the agents, a model is suggested for absorbing "shockwaves" of the dynamics even with individuals having only local information. The objective is to investigate the potential of empowerment-controlled agents, also in conjunction with centralized information provisions or external rules, to provide such intrinsic buffers as a model for urban transport of increased robustness and resilience.

While *empowerment* measures "freedom-of-operation", *sustainability* models its reversible part, the ability to undo when facing unintended consequences (Kim & Polani, 2009; 2021).

**Informational Models of Sustainability (sustainable empowerment models).** Sustainability is developed based on empowerment to model urban mobility considering on two different levels. Sustainability in the first sense conveys information about the ability of the system to recover from particular decision/movement patterns made in a particular situation, i.e., the level of local dynamics as reversibility of decision-making (measured by a "sustainability" version of the empowerment measure) in a spectrum of relevant scenarios. The second model of sustainability characterizes the amount of "commitment" of inducing particular topological or logistical changes, i.e., irreversibility, or the complement to "commitment" to a particular choice of topology, connectivity or regulation. Informed also by the "Empowerment-Based Dynamics" tools, sustainable empowerment permits the evaluation of "unintended consequences" of particular choices of topology or regulations and the degree to which they may be undone. The two models for the level of topological modifications of the urban structure provide a short-term scale of individual decision-making as well as mid-to-long term scale of topological/regulatory changes of the urban environment.

The empowerment-based informational framework and its cognitive-constraints connect the practical and the ethical aspects of city life and environmental values. An envisioned MaaS platform is expected to be capable of delivering services in a way that accounts for inclusion, cognition and information load when making daily decisions. Empowerment is a method applied to capture the freedom of choice, and other social and behavioral aspects of digital life in cities for better design, ontology, and civic mindedness (Baskerville, Myers, & Yoo, 2019).

**Goals and scenarios**

The informational framework now expands according to the questions that follow the proposed assumptions and principles. Tables 1-3 summarize the exploration of goals and scenarios that extend existing models of mobility services and service design, and the suggested service model of shared mobility, in the sharing economy.

**What should services incorporate? Urban cognition perspective**

Changing our approach, from silos to a holistic view, the scenarios and goals compose a new bricolage. The goal is to utilize resources and to gain the effective time and money trade-off, but not only. Utilizing technology for the benefit of other social groups and society at large, and for the livability of nature in the built environment, is also a possible system goal. What should a system of services incorporate? The multiple aspects considered when deciding and carrying out spatial behaviors may include ease of route planning, usefulness of navigation applications, reliability of service providers, quality of mobility services in general, familiarity with urban form, social aversion, perceived safety, tolerance to uncertainty, and more. Among the many components that shape a decision, some are more stable than others, e.g., time constraints can vary along the day; before and after check-in to a flight; or when facing competing goals that alter preferences and habits. Disabilities, on the other hand, impose a relatively stable constraint that limits the ability to reshape mobility decisions.

The ability to redefine decision components and to reshape previous mobility tendencies is, therefore, a power to control and shape one's mobility behaviors. Empowerment is indeed generally defined as the freedom of agents to determine their or their fellow agent's futures (Salge & Polani, 2017).

Furthermore, as the term implies, empowerment is also about the freedom of agents to priorities values, decision goals, and decision components. Empowered agents are free to move *and* to change their mobility goals to further increase their freedom – through flexibility in the regulation of their own movement in a way that improves coordination with other agents.

The freedom of agents to alter mobility preferences aims at liberating free energy rather than minimizing information load. The flexibility of their own movement – i.e., adaptation to the constantly (but systematically) changing environment, and self-regulation that enhances coordination (rather than the enforcement of top-down governance), make it possible to reduce the cognitive load involved in compliance to the control over their movement.

While both – the freedom to move as well as constraints on the freedom to move – are able to increase information load (either overflow, facing too many options, or the need to suitably comply, accommodate the appropriate response under uncertainty), there is a significant difference in the type of attention in demand. When creating internal representations of mobility decisions, the ability to choose the possible future is expected to preserve energy, free choice rather than complying to particular constraints. In other words, in addition to the *information* load (due to the need to properly follow specific procedures or arrangements), another type of attention demand is considered within the *cognitive* framework, carrying a *semantic* weight, related to the meaning (purpose and implications) of mobility decisions and actions. The semantic load may work in the opposite direction, rescuing cognitive energy, depending on its relation to metacognition.

Accordingly, the value of the mobility service is equivalent to the sum of information load, which holds a negative value, and semantic gain which is presumably a positive underlying metacognition, a mechanisms of information processing enhancement and propagation (Said, Fischer & Anders, 2020; Schwarz et al., 2021) throughout using the mobility service.

a. Information load

**Information in relation to the agent.** Different agents may experience the routes differently: exploration can be very easily described in informational terms, e.g., through the choice of the particular "quadrant" of interest. In principle, this is a discussion about the "boundary conditions/constraints" for the system which will determine the choice. Time constraint is a common condition (to avoid the stressful ride or to enjoy the challenge), but preferences may change over time, e.g., to "exploit" a favorite place or care for finding out whether there is something else to do, someone to assist; whether the trip is efficient or pleasant, an inspiring route to start your day off, or the opposite in the evening). These mental processes are complicated. The possible "operational clusters" means, which combinations make sense and what would drive an agent into one of the main combinations of drives; the choice of trade-off points is critical for the success, having both exploration and exploitation operate at a maximum. The need to balance is basically a meta-behavior, e.g., a choice in finding out about a game (exploration) or capturing resources from other competing agents (exploitation). Touristic exploration, high interest in cognitive absorption, may compete with a desire to exploit every minute. City exploration is usually not that critical, so we have significant more leeway to work with. We are usually dealing with the limit cases, namely people working in the city, not caring to explore, and just minimizing effort of getting from A to B or tourists who actually want to find out the salient points in the city. If they have a book or google maps, they will probably use that, but some tourists are good self-explorers, but this works best in a city with interesting landmarks (e.g., typical old European cities, where the build of the city itself "sucks" visitors in).

The objectification of everyday life in cities full of smart autonomous things deserves a new conceptualization, designed to meaningfully engage users (either gamification or through participatory planning) or solve the urgent problem of traffic congestion. The informational framework makes it clear that the efficiency of the (semi)automated routines is actually efficiency by rigidity and determinism (opposite to empowerment) that force the ritualized patterns, achieving coordinated behavior through constraints (costs bound behaviors under limited attention/memory). The cognitive load that decreases the capacity to calculate decreases freedom; in the absence of free choice, collective behavior is embodied.

**b. Urban morphology**

**Information in relation to the environment.** The spatial properties of the city (structural, functional) have an impact on the sensory system, perceptions and cognitive load. The models in the literature often combine distal efficiency (i.e., minimum distance) with a compromise on simplicity; exploration may be one desirable aspect, but having good access is another, such as optimal placing of fire stations. Possible scenarios and service goals are:

- Scenarios:
    - Periphery to inner city
    - Traffic jam in city. Empowerment-based car behavior (10% increase in flow).
    - Routine (every evening); one-time (to airport). Seasonal.
- Goals:
    - Optimization of the distance run between two points
    - Simplification of the walk between two points (relevant information minimization)
    - Robustness – which paths are less disruptible
    - Maximization of the possibilities to get anywhere (empowerment; fire brigade model)
    - Maximization of the exploration of your area

Some of these may partly or wholly coincide.

A universal level of view is set for the informational conditions: The cost is always an interplay between agent and environment. If the agent is limited, it influences what is easy and what is difficult to do. The choice of what the agent picks – making it subjective – is more difficult than modeling the whole trade-off curve, where we trade off efficiency for information processing effort – using an evolutionary argument, and the particular evolutionary niche sets the hypothesized trade-off curve. For the present discussion, the choice of the trade-off point is considered a property of the agent (e.g., interested in the shortest/fastest/most detailed path), as well as that of the road (highways vs. alleys), being a "niche negotiation". The travel cost between two points is a property of the world, and the complexity of the travel is a property of both, agent and world, intertwined.

The following section explains how an emotional component congregates with the informational framework to contribute semantic weights, much as values and preferences do, through urban partnerships that change power relations and economic arrangements.

**c. Semantic gain to balance information load**

Discrimination and the aversion from social interaction in ridesharing are among the implications reviewed above. An emotional level is assumed within the informational framework of urban reactions and behaviors, much as cultural values, that sustain caring for places, in addition to the spatial cognition, much as caring for people is a cultural expression in local communities, or "intimate anonymity" instead of close relationships among city dwellers. Cultural values of mutuality and reciprocity, or responsibility for others, is expected to fill the emotional layer in the informational framework and to ease the burden of cognitive/emotional overload. A sense of responsibility and locus of control, by caring for others, is then assumed to provide a relief, working through the moral-ethical *psychological* action of the person involved (that implicitly bounds the indifference and despair), aligning with a need to respond, design, make a difference. In the context of the informational framework, the compassionate attitude with its values and preferences can be referred to as a semantic level. Drawing on empowerment, emancipation (of the users) is declared in the framework of citizen science, not to "crowd-source" the users but to emancipate them, e.g., enriching their mobility capital – ability to move in various modes of transport, empower by altruistic choice; engage in rich system use (Burton-Jones & Straub, 2006), allowing agents with redefining their utility and connection to other users and systems (Al-Natour & Benbasat, 2009).

In the context of information theory, the goal of the proposed model, group-routine creation, would be to induce *phase transition* from social aversion to social inclusion (group), and from spontaneous individual ride hailing to reservation-based mobility service to stabilizes the overall traffic system-wealth (routine), i.e., from information load to semantic gain.

New ways to share data allow decoupling applications from the personal data they produce, which moves away from vendor lock-in to data ownership (e.g., Solid). The universal standards and open APIs has made interoperability possible across regions, vendors, and MaaS providers (ITF & OECD, 2021). An open, decentralized and universal structure is expected to produce a new value, termed *sustainability*, which is modelled using the approach of *reversible* empowerment dynamics (Kim & Polani, 2021; Salge, Glackin & Polani, 2014; Salge & Polani, 2017).

**d. Phase transition – obtaining groups and routines that are supply-oriented**

In this governance of many-stakeholders the supply-side and the demand-side are mutually interacting in a way that constantly changes representations and perceptions, attitudes and subjective norms – and the resulting expectations – to collaboratively induce *transitions*. Accordingly, the efficient allocation of mobility resources (utilizing available vehicles and free roads) and the empowerment of service users with education, are also expected to promote mobility capital in society, *motility* as proposed by Shliselberg & Givoni (2018).

However, traffic load in cities challenges the viability of current mobility paradigms to provide effective mobility services, as presented in Table 1. For example, the cost of multimodal public transport (time, planning effort), requires a complementary service, mostly in and across regions. Furthermore, the viability of current (TNC) and future (SAV) on-demand services is questioned, as shown in Table 2 in relation to price and time, and in Table 3 for flexible and reliable services.

A supply-based paradigm can therefore be proposed here to balance on-demand services that dominate the market of MaaS and lock shared mobility services in shared traffic jams.

The intention is not only reservation of travels in timetables but scheduling travels along the day considering time of day, places, passengers etc. Daily routines and other patterns of spatial behaviors can outline a fundamental demand layer in a supply-based paradigm that captures urban dynamics and spatial-temporal behaviors, mostly in groups, to stabilize activities that induce traffic loads. Therefore, we propose a switch from immediate response to scheduled supply that is *aware of demand but not on-demand*. In addition, public participation is part of the service design (as mentioned in Table1), to provide affordable mobility, and easy to use interfaces that improve accessibility to mobility services. This aligns with the need to switch from real-time requests, and the instant occurrence of responses that persists to rapidly reward unplanned actions, almost as an automatic result of stimulus-response learning, to planned behaviors that accounts for service coordination, preceded by consciously developed intentions, thus regulating goal-directed habitual behaviors (Kruglanski & Szumowska, 2020; Kurz, Gardner, Verplanken, & Abraham, 2015) in different frequencies, i.e., daily (commuting), weekly, or rare but recurrent (Schläpfer et al., 2021).

*Table 1. **Service models** affect the system, lead to a set of service sub-goals*

| Service model | Determining bound* and *service goal* | Cost of scaling | Possible solutions |
|---|---|---|---|
| **Public transport** (PT); based on travel surveys. Vehicles: dedicated bus; minibus | Routes; frequencies. *Service is **system**-oriented* | Cognitive load: planning effort of route, interchanges, first/last mile | Improve multimodality, synchronize interchanges. |
| **TNC, taxi**; real-time requests. Vehicles: private car | Add vehicles; build roads *Service is **user**-centred* | Congestion: social, economic, and environmental costs | Alternative for limited supply of roads: flying taxis; drones (energy waste of first order). |
| **Bubble**; real-time requests. Vehicles: dedicated minibus | Detours *Service is **users**-centred* | Cognitive load: uncertain availability, waiting and travel time | Empty rides (larger flee size for rush hours). |
| Proposed service: **Demand-supply coordinated** model ||||
| **Scheduled supply**; reservation group ridesharing. Vehicle: any available vehicle | ** Low demand hours and low access. areas. *Service is **socially** oriented* | Cognitive load (reservation ahead). | *** ICT utilization. |

\* Determining bound: what are we going to increase if demand increases (Duranton & Turner, 2011).
\*\* Demand in low-demand hours; and in less accessible areas: The supply of coordinated mobility services to new destinations/in new routes is expected to increase accessibility to new activities (as opposed to the problem of induced demand for private car when increasing the supply of roads).
\*\*\* ICT to assist: 1) Public participation to improve affordable mobility; 2) User interface for ease of use; 3) Utilizing available vehicles and free roads; 4) Empowerment with education towards *motility*.

While Table 1 frames concerns in mobility system design, Tables 2-3 show how different mobility systems address the main considerations of users and other stakeholders (Wolf, 2009).

*Table 2. **Price and speed in mobility services***

| High price (unclear, unstable) | | | TNC, taxi (a risk of slower and more expensive as of congestion) |
|---|---|---|---|
| Low price (fixed, relatively simple) | PT (a risk of higher price as of loss of travels to TNC in high demand city centres); Carpool; Bubble (slower as of detours) || **PT on-reservation:** public but flexible; known group (private but open) |
| | Low speed || High (flexible) speed |

*Table 3. **Service flexibility** (in time and route) and **certainty** (in reliability of service: availability, accuracy in arrival time; and uncertainty regarding social interaction in TNC and taxi services).*

| | | |
|---|---|---|
| **High certainty** | PT (fixed schedule and routes) | **PT on-reservation:** public but flexible; known group (private but open) |
| | Carpool (depends on others, can't book) | TNC, Taxi: lower if unavailable; can't book ahead; aversion to social interaction; driver/rider discrimination |
| **Low certainty** | Bubble (heigh variance in waiting time\ unavailability; and in travel time as of detours; no spontaneous ride-hailing). Flexibility is limited (detours; limited operation areas; bound to bus stops) | ⬇ |
| | Low flexibility | High flexibility |

To summarize, the current paradigm of on-demand services lacks attention to the supply-side of system resources, their renewability, and the overall system welfare. The absence of *supply*-oriented services that care for the *system* has a cost. In this work, caring for system resources is proposed to facilitate the cognitive effort: to minimize uncertainty, and to induce a cognitive gain, thus balance information load.

## 4. Discussion

On-demand mobility services with shared autonomous vehicles are expected to increase traffic congestion and unequal access to transport services. In this work we review mobility models and consolidate informational frameworks to approach the smartification of mobility services in cities and regions. Rethinking service design mechanisms in the shared economy, we wish to integrate the multiple aspects and trade-offs in mobility decisions and spatial behaviors into the theory of urban cognition. In order to develop theory-based *group ridesharing* service modelling, the paper reviews service design in mobility from different perspectives to explore and integrate design objectives:

**Coordination.** To conceptualize the phenomenon of mobility flows with individualized on-demand volatility vs. a coordinated group ridesharing, *thereby providing a mobility mechanism to stabilize traffic and improve intermodal services by synchronizing rides;* and theorize the MaaS community-building based on modeling of the group ridesharing behaviors with the goals to reduce excessive rides. Moreover, the proposed approach can enhance accessibility and freedom to move, and to deliver equitable access to diverse communities (Fagnant & Kockelman, 2015), and to vulnerable and undocumented users (World Bank, 2020) who cannot afford to reduce their mobility footprint. This approach can further encourage walkability in green spaces (e.g., ITF's Urban Passenger Transport Model 2020, p. 48 in OECD, 2021) that contribute to sustainable behaviors and happiness (e.g., Barrera-Hernández et al., 2020), and to embrace human-scale planning.

**Urban cognition.** The impact of urban morphology on spatial cognition and behaviors is considered to reduce the environmental, social and economic costs of traffic inefficiencies. Moreover, by reframing well-studied concepts such as gain-based behavior and rational decision making in the urban context, that must address dwellers' expectations, values and preferences, a hypothesis about the social and the environmental value of actions is made. According to the hypothesis, when considering cognitive load, the need to gain meaning possesses an emotional load that may add to the information load or relief as of its positive effect. Thus, in addition to strict constraints on the calculation of information load (e.g., minimizing uncertainty), cognitive load is a matter of choice, and mobility decisions that expand self-interests to include an altruistic perspective and prosocial behaviors has a semantic weight (metacognition) that, when positive, empowers the agent.

**Social action.** Gender and age differences, ethnic bias, and many other sources of disparity have social implications but, in turn, hold a design opportunity for new operative rationales in mobility. Mobility services are prone to system escalation and failure in cases where service is prevented due to discrimination. To redefine and resolve the boundaries among self-interested agents and social groups, a dynamic service should address the so-called taste variation (Vosooghi et al., 2019) and balance user differentiations in SAV demand modeling. In this paper social disparities and discrimination in mobility are introduced, to get a better understanding of social perceptions that suggest possible pro-social metacognitions.

Mobility services that draw on riders' choices to stabilize traffic and to reduce the environmental, social and economic costs of inefficient urban traffic, especially those relating to social polarization vs. cohesion of different social groups, to resolve the anomaly of *service providers* who dominate traffic and *service consumers* who comply with those market leaders' dominance although their size affords great market power (ITF & OECD, 2021).

The smart, autonomous, electric and shared transportation revolution is a promise waiting to be fulfilled. How do we track demand, tap into basic needs of travelers, and reduce traffic volumes? How do we best organize partnerships to *improve intermodal services* and *synchronize continues rides*? While this paper is focused on the reasoning towards mobility coordination, more work is required to scrutinize the underlying assumptions that derive the guiding principles in current mobility services, rethinking group ridesharing for community building.

A novel *informational framework* is proposed here in the context of smart mobility. The *informational framework* enables approaching empowerment and sustainability, as developed in this paper, to provide mobility that is aware of demand but not on-demand, thus designing a mobility service system of scheduled supply. The new approach is presented in this paper to address the anomaly in the current market-driven smartification of services, to justify partnership models; and to provide an *open* service design. More work is needed to better understand and design sharing mechanisms and to improve overall system overall system welfare in the service economy.